\documentclass[a4paper,11pt]{article}
\usepackage{graphicx,amssymb,amsmath,bm,latexsym,color,epsf,cite,comment}
\makeatletter
\@addtoreset{equation}{section}

\makeatother
\newcommand{\bea}{\begin{eqnarray}}
\newcommand{\eea}{\end{eqnarray}}
\newcommand{\vs}[1]{\vspace{#1 mm}}
\newcommand{\hs}[1]{\hspace{#1 mm}}

\renewcommand{\b}{\beta}

\newcommand{\G}{\Gamma}

\newcommand{\la}{\lambda}

\newcommand{\nn}{\nonumber\\}

\newcommand{\Tr}{{\rm Tr}}

\begin{document}


\begin{center}
{\Large\bf Wave Function Renormalization in Asymptotically Safe Quantum Gravity}
\vs{10}

{\large
Hikaru Kawai$^{a,b,}$\footnote{e-mail address: hikarukawai@phys.ntu.edu.tw}
and
Nobuyoshi Ohta$^{c,d,}$\footnote{e-mail address: ohtan.gm@gmail.com}
} \\
\vs{5}

$^a${\em Department of Physics and Center for Theoretical Physics, National Taiwan University, Taipei 106, Taiwan}

$^b${\em Physics Division, National Center for Theoretical Sciences, Taipei 106, Taiwan}
\vs{2}

$^c${\em Research Institute for Science and Technology, Kindai University, Higashi-Osaka, \\
Osaka 577-8502, Japan}

$^d${\em Nambu Yoichiro Institute of Physics, Osaka Metropolitan University,
Osaka 558-5585, Japan}

\vs{10}
{\bf Abstract}
\end{center}

We discuss the effect of wave function renormalization (WFR) in asymptotically safe gravity.
We show that there are two WFR-invariant quantities, and the renormalization (RG) equations may be written entirely
in terms of these quantities. The same set of RG equations can be obtained whether we fix the vacuum energy or Newton
coupling along the RG trajectory. The flow of the Newton constant and the vacuum energy is also discussed in detail.
In particular we discuss how the vacuum energy behaves near the singular barrier in the low energy.

\vs{10}
\setcounter{footnote}{0}

\section{Introduction}

The possibility of formulating quantum gravity in the context of ordinary field theory attracts
a lot of attention.
After the idea of the asymptotic safety was initiated in~\cite{Weinberg}, many considerations have been made
along this line. 
There have been several active attempts~\cite{Reuter,NR,perbook,rsbook}, based on the nonperturbative functional
renormalization group (RG) equation~\cite{Wetterich,Morris} to seek for asymptotically safe unitary theory.
The RG equation is given in terms of the  effective average action $\G_k$ as
\bea
\frac{d}{dt}\G_k=\frac12 \Tr \left(\G_k^{(2)}+R_k\right)^{-1}\frac{d}{dt} R_k,
\label{flow}
\eea
where $k$ is the cutoff, $R_k$ is a cutoff function suppressing the contribution of the modes below
the momentum scale $k$ to $\G_k$, 
$t=\log (k/k_0)$, $k_0$ being
an arbitrary initial value, and $\G_k^{(2)}$ is the second variation of the effective average action.

The RG equation~\eqref{flow} gives flows in the infinite-dimensional theory space.
However, in order to solve the flow equation~\eqref{flow}, we have to make approximation by restricting
this infinite-dimensional theory space to a finite subspace. This approximation is called truncation.
The approach of asymptotic safety attempts to find nontrivial fixed points away from asymptotically free theory
for a given truncation.
Once such fixed points are found, we have to extend the restricted space to study whether additional operators
are necessary to define the ultraviolet (UV) theory until there are no more necessary operators.

In general, any field theory contains redundant operators whose  coupling parameters change when we subject
the fields to a point transformation and have no effects on physical quantities.
It is well known that they need to be treated properly in order to study
fixed point actions. In quantum gravity, the couplings of such redundant operators are sometimes referred to as
inessential couplings~\cite{Weinberg}. The most typical of these is the wave function renormalization (WFR).
This appears in any truncation of the action because it preserves the form of each term.
In the perturbative approach, this point was discussed in Ref.~\cite{FT}, and in the context of asymptotic safety
in Ref.~\cite{KO}. Related field redefinition is also considered in \cite{BF,Knorr}.
The discussions of parametrization of the metric were made in \cite{GKL,OPP1,OPP2,GPS}.
The results when the WFR is taken into account are discussed in relation to causal dynamical
triangulation~\cite{Ambjorn1,Ambjorn2}. Here we further elaborate on the discussion.

In ordinary field theory, we usually fix the coefficient of the kinetic term by the WFR.
Unless the coupling of some operator is fixed, the theories on the RG trajectory cannot be compared.
Any one of the coupling constants, however, can be fixed, except those that are unchanged by
the field rescaling. In the case of quantum gravity, which consists of the Einstein and cosmological constant terms,
either the Newton coupling or the cosmological constant can be fixed using WFR.
It does not make physical sense to consider the flow of these couplings separately.
The choice of a parameter to be fixed is not important because the physics depends only on the combination
of the coupling constants which does not depend on the WFR.
For example, if the cosmological constant is fixed, the RG equation tells us the flow of the Newton coupling.
In our previous paper~\cite{KO}, we have shown that the RG equation can be written in terms of WFR-invariant quantity.
Here we show that there are two WFR-invariant quantities, and the RG equation may be written entirely
in terms of these quantities, thus the equations are manifestly invariant under the WFR.
We also show that the same set of equations can be obtained whether we fix the vacuum energy or Newton coupling
along the RG trajectory. The low-energy behavior of the vacuum energy is also discussed.

This paper is organized as follows. In sect.~\ref{Ein}, we start with the derivation of the RG equations
for the Einstein theory with the cosmological constant term.
In sect.~\ref{sec:wfr}, we consider the effect of WFR in this theory.
In subsect.~\ref{general}, we give general consideration of RG equations and their fixed points.
In subsect.~\ref{fixve}, we study the RG equations in more details, and find that they are
entirely written in terms of WFR-invariant quantities by fixing the vacuum energy along the RG trajectory.
As discussed in subsect.~\ref{general}, the same RG equations for WFR-invariant quantities should be
obtained if other quantity is fixed along the RG trajectory.
In subsect.~\ref{fixnc}, this is confirmed explicitly for the case when the Newton coupling is fixed.
In subsect.~\ref{ir}, we discuss the flow of the Newton coupling and the vacuum energy or
equivalently cosmological constant.
There has been argument that there is a singularity barrier in the low-energy for the cosmological constant
near $\lambda=1/2$ where $\lambda$ is the dimensionless cosmological constant~\cite{RS,CPR}.
We point out that this is not really the singularity but flow of the cosmological constant simply stops there.
Section~\ref{discussions} is devoted to summary and discussions.

\section{Einstein gravity with the cosmological constant}
\label{Ein}

Following the notation of \cite{Weinberg1},
we make slight change of notation compared with \cite{KO}:
\bea
S = \int  d^4 x \sqrt{g}\, \left( \rho- \frac{1}{16\pi G} R \right),
\label{bareaction}
\eea
where $G$ is the Newton coupling constant and $\rho$ is the vacuum energy.
The cosmological constant $\Lambda$ appearing in the Einstein equation is related to $\rho$ by
\bea
\rho =\frac{\Lambda}{8 \pi G}.
\label{stL}
\eea

Using the standard cutoff
\bea
R_k=(k^2-\Delta) \theta(k^2-\Delta); \qquad
\Delta \equiv -g^{\mu\nu}\nabla_\mu \nabla_\nu,
\label{cutoff_s}
\eea
in the RG equation given in~\eqref{flow}, we get the RGE for the coefficients as~\cite{CPR}
\bea
\frac{d\rho}{dt} &=& \frac{k^4}{16\pi}\left(A_1 + G \left(\frac{d}{dt}\frac{1}{G}\right) A_2 \right), \nn
-\frac{d}{dt}\left(\frac{1}{G}\right) &=& k^2 \left(B_1 + G \left(\frac{d}{dt}\frac{1}{G}\right) B_2 \right),
\label{beta1}
\eea
where
\bea
&& A_1 = \frac{1+64\pi G \rho k^{-2}}{\pi(1-16\pi G \rho k^{-2})},\qquad
A_2 = \frac{5}{6\pi(1-16\pi G \rho k^{-2})},\nn
&& B_1 = -\frac{11-144\pi G \rho k^{-2} +7(16\pi G \rho k^{-2})^2}{3\pi(1-16\pi G \rho k^{-2})^2},
\qquad
B_2 = -\frac{1+80\pi G \rho k^{-2}}{12\pi(1-16\pi G \rho k^{-2})^2}.
\label{coef1}
\eea
We can solve Eqs.~\eqref{beta1} for $\frac{d\rho}{dt}$ and $\frac{d}{dt}(\frac{1}{G})$ to obtain
\bea
&& \frac{d\sqrt{\rho}}{dt} = \frac{k^4}{16\pi} \left(A_1-\frac{k^2 G A_2 B_1}{1+k^2 G B_2}\right)
\frac{1}{2\sqrt{\rho}} \equiv \b^1, \nn
&& \frac{d}{dt}\left(\frac{1}{G}\right) = - k^2 \frac{B_1}{1+k^2 G B_2}
\equiv \b^2,
\label{frge}
\eea
where we have defined the beta functions $\b^1$ for $\sqrt{\rho}$ and $\b^2$ for $1/G$ for later convenience.

\section{Wave function renormalization and renormalization group}
\label{sec:wfr}

As discussed in \cite{KO}, the WFR
\bea
g_{\mu\nu} =Z g_{\mu\nu}',
\label{wfr1}
\eea
causes the change of the couplings as
\bea
\rho' = Z^2 \rho, \qquad
G' = Z^{-1} G.
\eea

\subsection{General consideration}
\label{general}

Let us first make general consideration.
For this purpose, it is convenient to set the mass dimension of all the couplings to 2.
Namely we define
\bea
(k^2, \sqrt{\rho}\, ,G^{-1}) \equiv (X^0, X^i) \equiv X,~~~ (i=1,2).
\eea
The equivalence under the WFR means that the theory is equivalent under the transformation
\bea
X \sim \la X.
\label{wfr2}
\eea
Here and in what follows, $\sim$ means that the theory is equivalent.

On the other hand, RG flow gives
\bea
((1+d\tau)X^0, X^i+\b^i(X) d\tau) \sim (X^0, X^i).
\eea
Here we use $\tau=\log(k/k_0)^2 (=2t)$ for the convenience of our discussions.
This implies that
\bea
X^0 \frac{dX^i}{dX^0}=\b^i(X),
\label{rge1}
\eea
and invariance under the WFR~\eqref{wfr2} tells us that
\bea
\b^i(\la X)=\la \b^i(X).
\eea
We can rewrite~\eqref{rge1} as
\bea
\frac{d}{d\tau} X^0 &=& X^0, \nn
\frac{d}{d\tau} X^i &=& \b^i(X).
\eea
When the WFR is taken into account, we have
\bea
\frac{d}{d\tau} X^0 &=& X^0 +\zeta(X) X^0, \nn
\frac{d}{d\tau} X^i &=& \b^i(X) +\zeta(X) X^i.
\eea

Now let us introduce the WFR-invariant variables
\bea
\tilde X^i =\frac{X^i}{X^0},
\eea
and then we find
\bea
\frac{d}{d\tau} \tilde X^i &=& \b^i(1,\tilde X) -\tilde X^i.
\eea
The UV fixed point is given by
\bea
\b^i(1,\tilde X)-\tilde X^i =0.
\label{uvfp1}
\eea

Using the freedom from WFR, we can fix $X^1$, say, to $1$.
This leads to
\bea
\zeta(X)=-\b^1(X),
\eea
and the RG equations become
\bea
\frac{d}{d\tau} X^0 &=& X^0(1-\b^1(X)), \nn
\frac{d}{d\tau} X^i &=& \b^i(X) -\b^1(X) X^i,~~~ (i=2,3).
\eea
This means that the UV fixed point is given by
\bea
X^1=1,~~~
\b^1(X)-1=0, ~~~
\b^i(X)-\b^1(X) X^i =0,~~~(i=2,3).
\label{uvfp2}
\eea
Clearly this agrees with the fixed point given by \eqref{uvfp1}.
It is also obvious we get the same fixed points if we change the fixed quantity from $X^1$ to another $X^i$.

\subsection{Fixing the vacuum energy}
\label{fixve}

Let us now formulate the RGE in more detail.
To make the standard cutoff given in \eqref{cutoff_s} do not break the invariance
under the WFR~\eqref{wfr1}, we have to transform the momentum cutoff scale $k$ under the WFR as~\cite{KO}
\bea
{k'}^2= k^2 Z.
\eea
In this case, it is convenient to make WFR-invariant combination
\bea
\ell^2 \equiv \frac{k^2}{\sqrt{\rho}}.
\label{ell}
\eea
If we define the quantities $\bar\rho$ and $\bar G$ by
\bea
\rho = \ell^4 \bar\rho, \qquad
G= \ell^{-2} \bar G,
\label{fixing}
\eea
we can make $\bar\rho$ invariant along the RG trajectory.
Indeed, from the sequence of the change under the RG flow and WFR:
\bea
&& (k^2, \sqrt{\rho}, G^{-1})\ \stackrel{\rm RG}{\sim}\ ((1+ 2 dt)k^2, \sqrt{\rho}+\b^1 dt, G^{-1}+\b^2 dt) \nn
&& \ \stackrel{\rm RG+WFR}{\sim}\ ((1+2 dt+\zeta)k^2, \sqrt{\rho}+\b^1 dt+\zeta\sqrt{\rho}, G^{-1}
+\b^2 dt+\zeta G^{-1}),
\label{change}
\eea
where $\b^1$ and $\b^2$ are the beta functions defined in \eqref{frge},
and $\zeta$ represents the changes of the variables, the above requirement implies that
\bea
\bar\rho=\frac{\rho}{\ell^4}=\frac{\rho^2}{k^4},
\eea
should be invariant under \eqref{change}. This leads to the condition
\bea
\zeta= 2 \left(1- \frac{\b^1}{\sqrt{\rho}}\right) dt.
\eea
Taking this into account, we have
\bea
\left\{
\begin{array}{l}\vs{2}
dk = \left(2-\frac{\b^1}{\sqrt{\rho}}\right) k dt, \\ \vs{2}
d\sqrt{\rho} = \left(2-\frac{\b^1}{\sqrt{\rho}}\right) \sqrt{\rho}\, dt, \\
d G^{-1} = \left(\b^2 + 2\left(1- \frac{\b^1}{\sqrt{\rho}}\right) G^{-1} \right) dt .
\end{array}
\right.
\label{id1}
\eea
Now we define another WFR-invariant physical quantity introduced in \cite{KO}:
\bea
\eta = 16\pi G\sqrt{\rho}.
\label{inv2}
\eea
We then find
\bea
\frac{d\ell^2}{dt} &=& \left(2-\frac{\b^1}{\sqrt{\rho}} \right) \ell^2, \nn
\frac{d\eta}{dt} &=& -\frac{\b^2}{16\pi \sqrt{\rho}}\, \eta^2 + \frac{\b^1}{\sqrt{\rho}}\, \eta.
\label{invrg}
\eea

Substituting $\b^1$ and $\b^2$ from \eqref{frge}, we find
\bea
\label{rg1}
\frac{d\ell^2}{dt} &=& (2-g)\ell^2 \equiv \b_{\ell^2}, \\
\frac{d\eta}{dt} &=& g \eta +f \eta^2 \equiv \b_{\eta},
\label{rg2}
\eea
where
\bea
f &=& \ell^2 \frac{B_1}{16\pi+\ell^2 \eta B_2}
= - \frac{4 \ell^2 (11\ell^4-9 \ell^2 \eta+7 \eta^2)}
{192\pi^2 (\ell^2-\eta)^2-\ell^4 \eta(\ell^2+5\eta)} , \nn
g &=& \frac{\ell^4}{32\pi} \left(A_1- \eta A_2 f \right)
=\frac{\ell^4 [\ell^4(107 \ell^2-10\eta)\eta+576\pi^2(\ell^4+3\ell^2 \eta-4\eta^2)]}
{96\pi^2 [192\pi^2 (\ell^2-\eta)^2-\ell^4 \eta(\ell^2+5\eta)]}
.
\eea
The real fixed points of these equations are
\bea
(\eta_*, \ell^2_*) = \mbox{P}_1: (3.7065, 9.5923),\ \mbox{P}_2:(0, 8\pi).
\label{fp1}
\eea
These correspond to
\bea
(\tilde\rho_*, \tilde G_*) = \left( \ell_*^{-4}, \frac{\eta_* \ell_*^2}{16\pi}\right) =(0.01087, 0.7073),
\eea
in agreement with \cite{CPR}.
The eigenvalues of stability matrix are, in the above order, given by
\bea
&& \mbox{P}_1: -1.4753 \pm 3.0432i, \nn
&& \mbox{P}_2: -4, 2.
\label{sta1}
\eea

\begin{figure}[tb]
\begin{center}
\includegraphics[width=90mm]{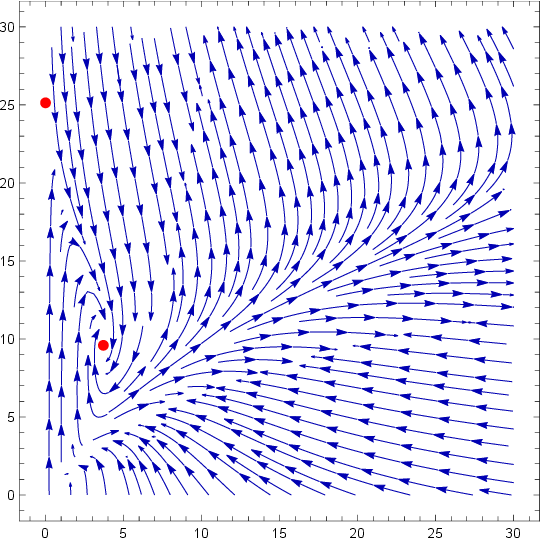}
\put(-270,230){$\ell^2$}
\put(10,0){\bf $\eta$}
\end{center}
\caption{RG flow in $\eta$-$\ell^2$ plane. The fixed points P$_1$ and P$_2$ are indicated by red bullets.}
\label{f1}
\end{figure}

In Fig.~\ref{f1}, we depict the RG flow and the fixed points.
The fixed point P$_1$ is an attractive point in the UV, and the fact that the eigenvalues of the stability
matrix is complex means that the flow is converging to this point with spiral curve.
The fixed point P$_2$ gives a separatrix.
This means that there is a trajectory into this point in the IR under the very
fine tuning, but the flow moves to either positive or negative infinity if
the initial condition is slightly away from it.
We shall discuss the structure of the flow in more detail in subsect.~\ref{ir}.

\subsection{Fixing the Newton coupling}
\label{fixnc}

We note that the above results are derived under the condition that $\bar\rho$ is invariant along the RG trajectory.
As we pointed out in subsect.~\ref{general}, the same set of equations should follow if we fix $\bar G$
along the RG flow. Here we show this explicitly.

We defined $\bar G$ in Eq.~\eqref{fixing} as
\bea
\bar G = \ell^2 G = \frac{k^2}{\sqrt{\rho}} G.
\eea
If we require that this be invariant along the RG trajectory, we find from \eqref{change} that this gives the condition
\bea
\zeta= \left( 2 - \frac{\b^1}{\sqrt{\rho}}-\b^2 G \right).
\eea
This leads to
\bea
\left\{
\begin{array}{l}\vs{2}
dk = \left(2-\frac{\b^1}{2 \sqrt{\rho}} -\frac{\b^2}{2} G \right) k dt, \\ \vs{2}
d\sqrt{\rho} = \left(2-\frac{\b^2}{2}\right) \sqrt{\rho}\, dt, \\
d G^{-1} = \left( 2 - \frac{\b^1}{\sqrt{\rho}}\right) G^{-1} dt .
\end{array}
\right.
\label{id2}
\eea
Using these relations with the definition of the invariant quantities~\eqref{ell} and \eqref{inv2},
we precisely reproduce the WFR-invariant RG~\eqref{invrg}.

\subsection{IR behavior}
\label{ir}

It is interesting to study what these RG equations tell us about low energy behavior of the Newton constant
and vacuum energy.

In order to see the IR behavior of the vacuum energy $\rho$, it is convenient to express the RG equation
in terms of two variables defined by
\bea
\left\{
\begin{array}{l}\vs{2}
\displaystyle{
x= \frac{1}{\pi}Gk^2}, \\
\displaystyle{
z= 16\pi^2\frac{\rho}{k^4}} .
\end{array}
\right.
\label{defxz}
\eea
Then the RG equation becomes
\bea
\left\{
\begin{array}{l}\vs{2}
\displaystyle{
\frac{dx}{dt}= \frac{2x}{1+b_2 x}\left\{1+(\frac{1}{2}b_1 +b_2) x \right\} \equiv \beta_x}, \\ \vs{2}
\displaystyle{
\frac{dz}{dt}= \left\{-4z+\frac{a_1+(a_1b_2-b_1a_2)x}{1+b_2 x} \right\} \equiv \beta_z},
\end{array}
\right.
\label{RGxz1}
\eea
where
\bea
\left\{
\begin{array}{l}\vs{2}
\displaystyle{
a_1= \frac{1+4xz}{1-xz}}, \\ \vs{2}
\displaystyle{
a_2= \frac{5}{6}\frac{1}{1-xz}}, \\ \vs{2}
\displaystyle{
b_1= -\frac{1}{3}\frac{11-9xz+7(xz)^2}{(1-xz)^2}}, \\
\displaystyle{
b_2= -\frac{1}{12}\frac{1+5xz}{(1-xz)^2}} .
\end{array}
\right. 
\label{RGxz2}
\eea

\begin{figure}[tb]
\begin{center}
\includegraphics[width=70mm]{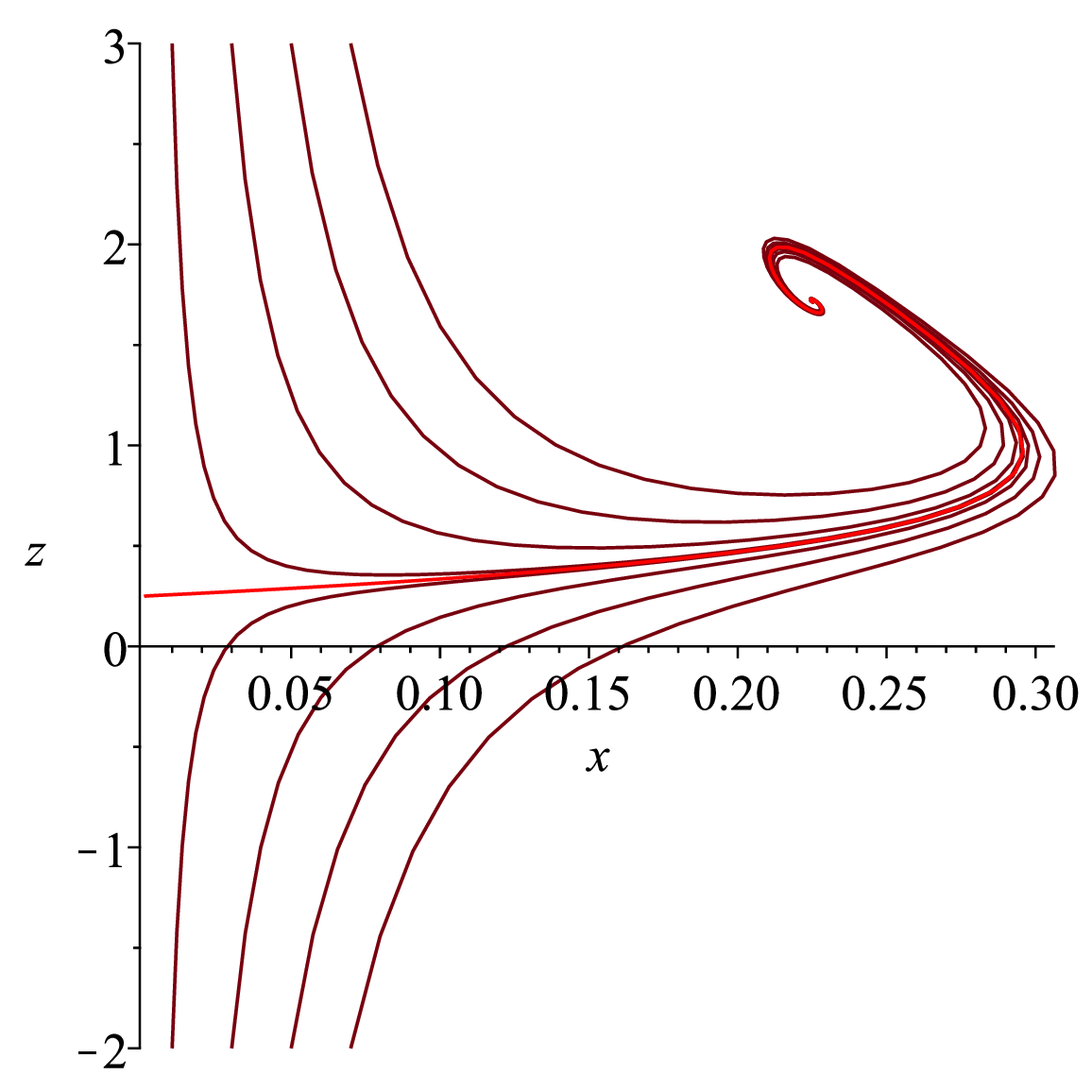}
\hs{10}
\end{center}
\vs{8}
\caption{The renormalization group flow in $x$-$z$ plane.}
\label{f2}
\end{figure}

In Fig.~\ref{f2}, we depict the RG flow in $x$-$z$ plane.
From this, we see that all the IR flows starting from near the UV fixed point 
get close to the $z$-axis ($x\sim0$). Depending on the initial value, $z$ diverges to either plus or minus
infinity when the flow approaches the $z$ axis.

Near the $z$-axis ($x\ll 1, xz\ll 1$), these equations become very simple:
\bea
\left\{
\begin{array}{l} \vs{2}
\displaystyle{
\beta_x\sim 2x }, \\
\displaystyle{
\beta_z\sim 1-4z } .
\end{array}
\right.
\label{RGIR}
\eea
From this, we have the approximate flow equation in the region $x\ll 1, xz\ll 1$,
\bea
\frac{dz}{dx}=\frac{1}{2x} \left(1-4z\right),
\label{flowxz}
\eea
which is solved as
\bea
z=\frac{1}{4}+\frac{c}{x^2},
\label{xzsol}
\eea
where $c$ is an integration constant.
This is consistent with the behavior of Fig.~\ref{f2}.
Note that $c=0$ corresponds to the red curve in Fig.~\ref{f2}.

The point $(x,z)=(0,\frac{1}{4})$ is an IR fixed point, which is nothing but P$_2$ already found in \eqref{fp1}.
The RG trajectory through it (the red curve in Fig.~\ref{f2})
corresponds to the case of zero physical vacuum energy. Here, the physical vacuum energy
means what is observed in a sufficiently large universe. In this sense the deviation from $z=\frac{1}{4}$ at $x=0$
(or the deviation from the red curve) gives the physical vacuum energy.
The explicit relation between the vacuum energy and the constant $c$ in \eqref{xzsol}
can be obtained as follows. 

By substituting \eqref{defxz} to \eqref{xzsol}, we have a relation that holds in the IR region  ($Gk^2\ll 1$, $G\rho k^{-2}\ll 1$):
\bea
c=16G^2(\rho-\frac{k^4}{64\pi^2}) .
\label{cGrho}
\eea 
This tells us that the zero physical vacuum energy ($c=0$) is realized when the bare vacuum energy is chosen to be
\bea
\rho_0=\frac{k^4}{64\pi^2} .
\label{rho0}
\eea 
Then by setting
\bea
\left\{
\begin{array}{l}\vs{2}
\rho= \rho_0+\rho_{\mathrm{phys}}, \\
G=G_{\mathrm{phys}}.
\label{low}
\end{array}
\right.
\label{rhoGphys}
\eea
\eqref{cGrho} becomes in the IR limit ($G_{\mathrm{phys}}k^2 \ll 1, G_{\mathrm{phys}}\rho_{\mathrm{phys}}k^{-2} \ll 1$)
\bea
c=16G_{\mathrm{phys}}^2\rho_{\mathrm{phys}} .
\label{cGrhophys}
\eea 
Note that $G_{\mathrm{phys}}$ and $\rho_{\mathrm{phys}}$ are the physical Newton coupling and vacuum energy
which are observed in a sufficiently large universe, and that the combination $G_{\mathrm{phys}}^2\rho_{\mathrm{phys}}$
is invariant under wave function renormalization. In the standard notation of the cosmological constant
defined as~\eqref{stL}, this gives
\bea
\Lambda_{\rm phys} = \frac{\pi c}{2 G_{\rm phys}}.
\eea
If $c=0$ by some mechanism, this gives vanishing cosmological constant.
We note again that the RG flow goes to the fixed point P$_2$ only for the zero cosmological constant.
Similar behavior of the cosmological constant $\Lambda(=8\pi G\rho) \sim (a+b k^2)$ was suggested in \cite{KS,PS}.
See also~\cite{CMO}.

\begin{figure}[tb]
\begin{center}
\includegraphics[width=70mm]{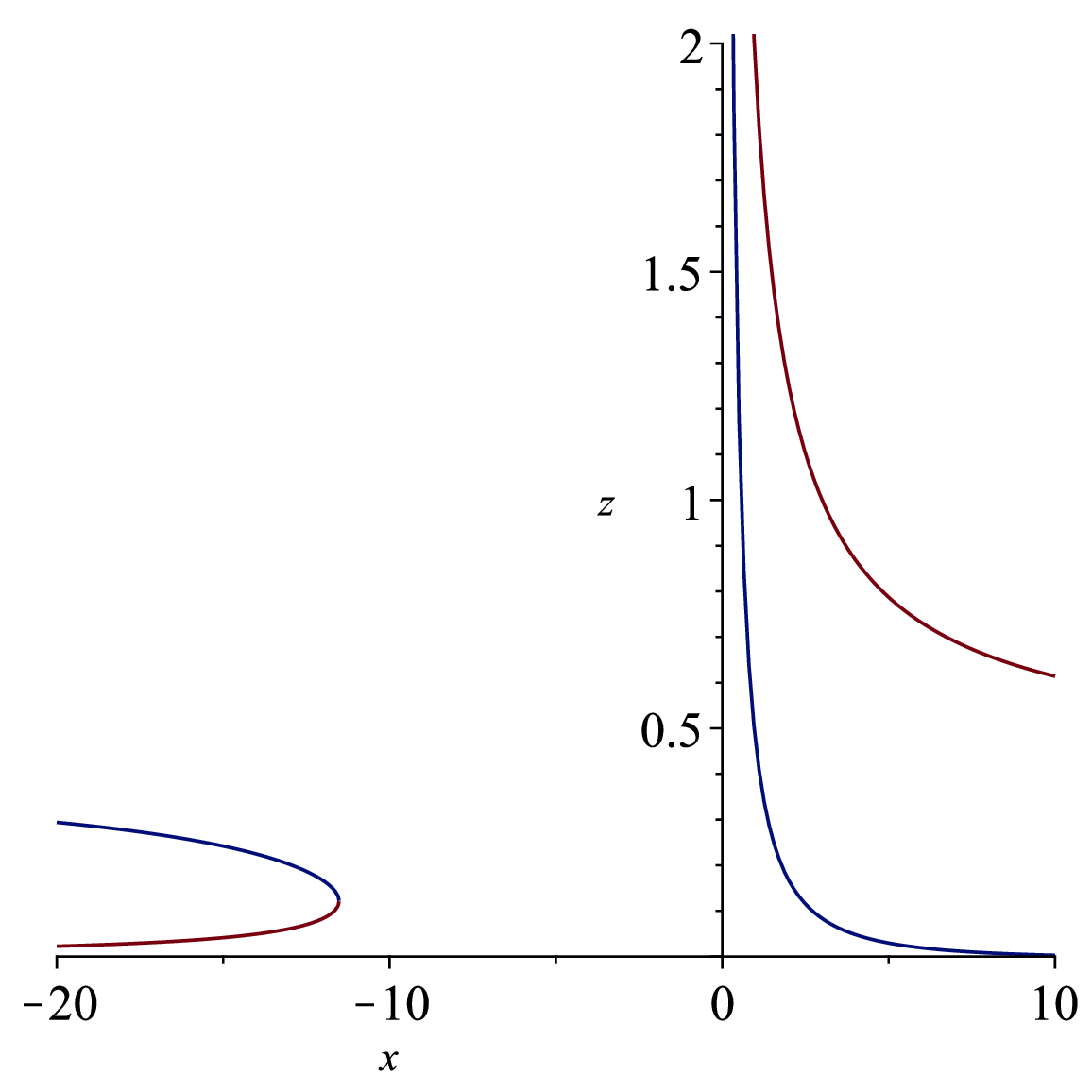}
\put(-185,-60){(a) Singularity of the beta functions}
\hs{10}
\includegraphics[width=70mm]{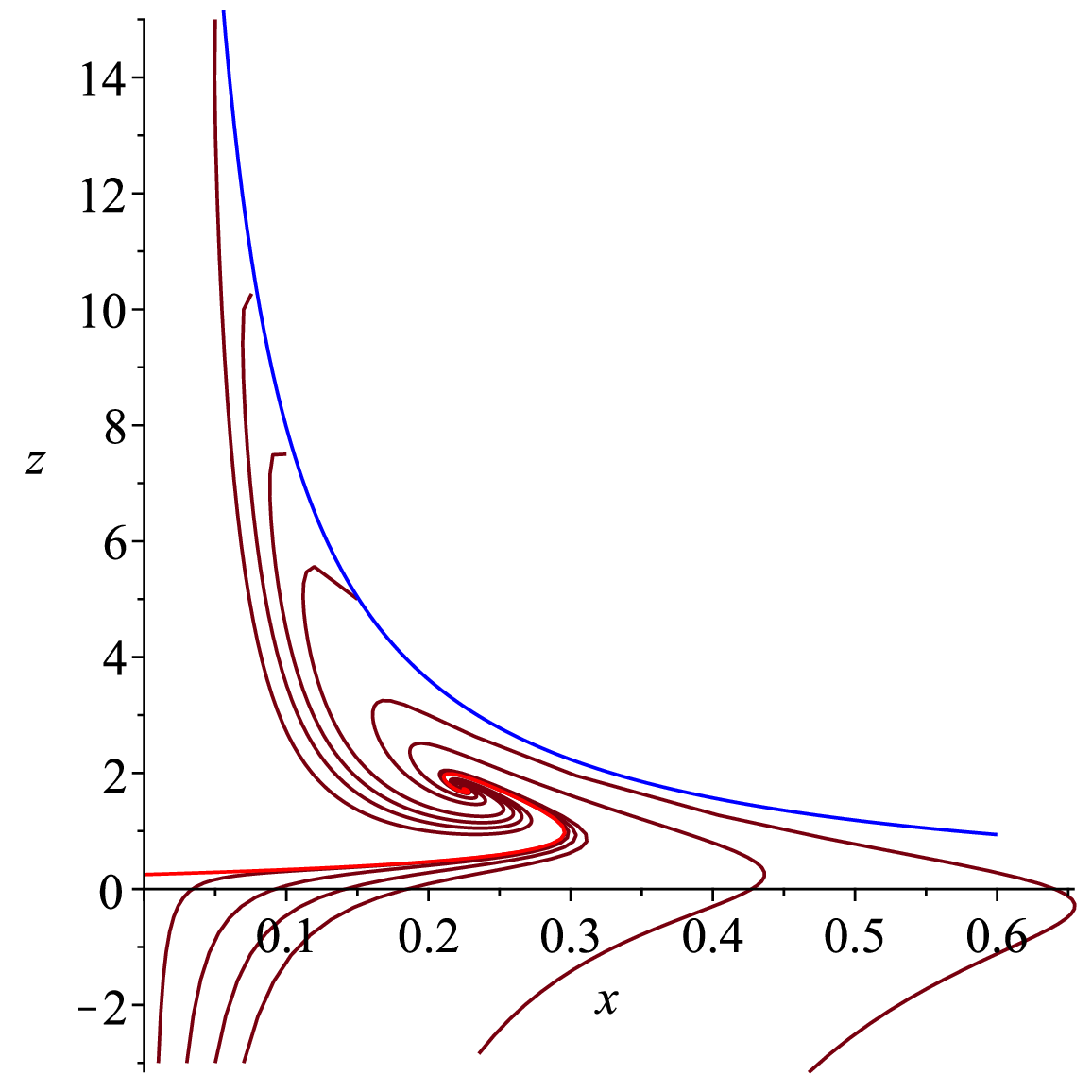}
\put(-170,-60){(b) RG flow and singularity}
\\
\end{center}
\caption{The blue curve in (b) is the lower curve in $x>0$ region in (a).}
\label{f3}
\end{figure}

We now discuss the singularity of the RG flow.
The beta functions given by ~\eqref{RGxz1} and ~\eqref{RGxz2} can be expressed as
\bea
\left\{
\begin{array}{l}\vs{2}
\displaystyle{\beta_x= \frac{a}{h}}, \\
\displaystyle{\beta_z=\frac{b}{h}} ,
\end{array}
\right.
\label{betaFactorized1} 
\eea
where
\bea
\left\{
\begin{array}{l}\vs{2}
a=2x [x(14x^2 z^2 -13 xz+23) -12(1-xz)^2] , \\
\displaystyle{
b=\frac{1}{3}[144(1-xz)^2 z +12 (7 x^2 z^2-10xz-3)+x (10 xz-107)]},  \vs{2} \\
h= 5x(xz+1)-12(1-xz)^2 . 
\end{array}
\right.
\label{betaFactorized2}
\eea
The zeros of $h$ are drawn in Fig.~\ref{f3}(a). The beta functions $\beta_x$ and $\beta_z$ go to infinity
simultaneously on these curves, so the one-loop approximation used here breaks down near these curves.
Fortunately, the region near the red curve in Fig.~\ref{f2} is away from the curve in Fig.~\ref{f3}(a),
so there is no need to worry about singularities, at least near the UV fixed point.
In fact, the lower curve in $x>0$ region in Fig.~\ref{f3}(a) is given by
\bea
z=\frac{1}{24x}\left(5x+24-\sqrt{25x^2+288x}\right) .
\label{singularity}
\eea 
For example the value of $z$ at $x=0.3$ is about $2.23$, which is about twice higher than the red curve
[see Fig.~\ref{f3}(b)].

However, in the IR limit the singularity becomes important if the physical vacuum energy is positive.
This can be easily seen from \eqref{xzsol} and \eqref{singularity}. Near the $z$-axis ($x\sim 0$),
\eqref{singularity} is approximated by
\bea
z=\frac{1}{x}+\mathcal{O}(x^{-1/2}) .
\label{singappr}
\eea 
Therefore if we see the RG trajectory from the UV side, it meets the singularity when 
$\frac{1}{4}+\frac{c}{x^2}\sim \frac{1}{x}$ becomes satisfied, that is,
\bea
x\sim c.
\label{EndFlow}
\eea
By substituting  \eqref{defxz} and \eqref{cGrhophys} to \eqref{EndFlow}, we have
\bea
k^2\sim 16\pi G_{\mathrm{phys}}\rho_{\mathrm{phys}} \sim 6H^2,
\label{kHubble}
\eea
where $H$ is the Hubble parameter.
This result is not surprising. This is because it merely states that the approximation used here is not applicable
at scales larger than the Hubble distance.

Finally, we make a small comment on the relation between the RG flow trajectory and the singularity.
From \eqref{betaFactorized1} we have the differential equation to determine the RG flow trajectory:
\bea
\frac{dz}{dx}=\frac{b}{a}.
\label{trajectory}
\eea
The important point is that $h$ cancels out from this equation, and thus the trajectory is determined
independently of the singularity at $h=0$.
In this sense, this is not a singularity in the equation determining the flow line.
This is the reason why there is no violent behavior in the flow line.
However, $h$ changes sign when it crosses the curve $h=0$. Therefore, the RG flow cannot cross the curve of
singular points, and the flow ends when the trajectory hits the singularity. As the flow approaches the singularity,
the $\beta$ function goes to infinity, so the time taken for the collision with the singularity is finite.
Therefore, the end of the flow occurs at a finite value of $t$ (or a finite value of $k$) and
the flow ceases to exist thereafter, as depicted in Fig.~\ref{f3}(b).

\section{Summary and discussions}
\label{discussions}

In this paper we have reformulated the RG equations by incorporating the effects of WFR.
We find that the RG equations can be entirely written in terms of WFR-invariant quantities.
It is quite reassuring that the resulting fixed points turned out to be the same as the analysis
without considering the effect, at least when the optimized cutoff is used~\cite{CPR}.

In our previous paper~\cite{KO}, we pointed out that the momentum scale $k^2$ should be scaled by
the WFR factor in order to keep the invariance of the RG equations. This point is correct, but
we fixed one of the WFR-invariant quantities $\ell^2=1$, which actually runs under the RG flow.
This led to different fixed point values. This arises because the fixed point value of $\ell^2$ takes not 1 but 9.5923,
as given by P$_1$  in \eqref{fp1}.

One could also introduce other WFR-invariant quantities
\bea
\tilde \ell^2 \equiv Gk^2, \qquad
\tilde \eta \equiv 16 \pi G^2 \rho,
\eea
and derive the RG equation by requiring that either
\bea
\tilde{\bar \rho} = \frac{\rho}{\tilde \ell^4}, \qquad
\mbox{ or } \qquad
\tilde{\bar G} = G \tilde\ell^2,
\eea
be fixed along the RG trajectory. Again both cases give the same set of RG equations
\bea
\frac{d \tilde\ell^2}{dt} &=& (2- \tilde g )\tilde\ell^2, \nn
\frac{d \tilde\eta}{dt} &=& \tilde f-2 \tilde g\tilde\eta,
\eea
where
\bea
\tilde g &=& \frac{4 \tilde\ell^2 (11 \tilde\ell^4 -9 \tilde\ell^2 \tilde\eta +7 \tilde\eta^2)}
{12\pi (\tilde\ell^2- \tilde\eta)^2 -\tilde\ell^4 (\tilde\ell^4 + 5\tilde\eta)},\nn
\tilde f &=& \frac{\tilde\ell^4 [107 \tilde\ell^6 +2 \tilde\ell^4(18\pi-5\tilde\eta)+108 \pi \tilde\ell^2 \tilde \eta
 -144 \pi \tilde\eta^2]}{3\pi [12\pi (\tilde\ell^2- \tilde\eta)^2 -\tilde\ell^4 (\tilde\ell^4 + 5\tilde\eta)]}.
\eea
We find a real fixed point
\bea
(\tilde\ell^2_*, \tilde\eta_*)=(0.7073,0.2733).
\eea
One can check that this precisely corresponds to the fixed point P$_1$ found before.
This formulation may appear preferable because $\sqrt{\rho}$ does not appear, but it seems difficult to find the second
fixed point P$_2$ in this formulation.

We have also studied the low-energy behavior of the vacuum energy.
The second fixed point P$_2$ in \eqref{fp1} actually corresponds to unstable fixed point in the low-energy.
Exactly on the line going into this fixed point, we have zero physical vacuum energy,
but just slightly away from this fixed point P$_2$, the vacuum energy flows to arbitrary values in opposite directions
with respect to this fixed point. In this sense, this is called separatrix.
There has been argument that there is a singular barrier near $xz=1$ (or near $\lambda=1/2$ for the dimensionless
cosmological constant) beyond which the vacuum energy cannot flow.
We have shown that this is not really singularity in the equation determining the flow line, but
the flow simply stops there.

If we could find some mechanism to choose the trajectory going into the fixed point P$_2$, we may have
vanishing vacuum energy in the low-energy limit, and the current tiny vacuum energy may be
explained within the FRG. Otherwise there remains an arbitrary value of the vacuum energy.
It is an interesting challenge to find such a mechanism.

\section*{Acknowledgments}

This work was motivated by the questions made by Jan Ambj{\o}rn whom we would like to thank.
H.K. thanks Prof. Shin-Nan Yang and his family for their kind support through the Chin-Yu
chair professorship. H.K. is partially supported by JSPS (Grants-in-Aid for Scientific Research
Grants No. 20K03970), by the Ministry of Science and Technology, R.O.C.
(MOST 111-2811-M-002-016), and by National Taiwan University.
The work of N.O. was supported in part by the Grant-in-Aid for Scientific Research Fund of the JSPS (C) No. 20K03980.

\end{document}